%% file: cmnMinFairness.tex
\documentclass[conference,romanappendices,hidelinks,10pt]{IEEEtran}
\IEEEoverridecommandlockouts
\IEEEoverridecommandlockouts
\usepackage{algorithm,algorithmic,amsmath,amssymb,amsthm,bbm,cite,color,graphicx,microtype,url}
\usepackage{tikz}
\usepackage[USenglish]{babel}
\usepackage[utf8]{inputenc}
\usepackage[T1]{fontenc}
\usepackage{comment}
\usepackage{epstopdf}
{}
{}
{}
{}
{}
{}
{}

\input{./Definitions.tex}
\newcommand{\bs}{\textnormal{\tiny{BS}}}
\newcommand{\dl}{\textnormal{\tiny{DL}}}
\newcommand{\dlA}{\textnormal{\tiny{DL-1}}}
\newcommand{\dlB}{\textnormal{\tiny{DL-2}}}

\newcommand{\sinr}{\mathrm{SINR}}
\newcommand{\ue}{\textnormal{\tiny{UE}}}
\newcommand{\ul}{\textnormal{\tiny{UL}}}
\newcommand{\ulA}{\textnormal{\tiny{UL-1}}}

\title{Joint Minimum DL-UL Rate Maximization \\ for Cell-Free Massive MIMO}
\vspace{-5mm}
\author{Bikshapathi Gouda, Italo Atzeni, and Antti Tölli \\
Centre for Wireless Communications, University of Oulu, Finland \\
Emails: \{bikshapathi.gouda, italo.atzeni, antti.tolli\}@oulu.fi
\thanks{\vspace{-5mm}

The work of B.~Gouda and A.~Tölli was supported by the Academy of Finland under grant no. 318927 (6Genesis Flagship). The work of I.~Atzeni was supported by the Marie Sk\l{}odowska-Curie Actions (MSCA-IF 897938 DELIGHT).} \vspace{-2mm}}

\begin{document}
\maketitle

\begin{abstract}
In cell-free massive multiple-input multiple-output (MIMO) systems, the beamforming strategies at the base stations (BSs) and user equipments (UEs) can be computed building on bi-directional training~\cite{Atz21,Gou20,Atz20}. However, the precoding/decoding optimization in the downlink (DL) and in the uplink (UL) generally requires two separate bi-directional training phases, which can be wasteful in the case of short scheduling blocks. This paper proposes a framework to reduce the bi-directional training overhead by considering a common beamforming training strategy for both DL and UL when the UEs to be served in the two directions are the same. In doing so, we consider the problem of maximizing the (weighted) minimum DL-UL rate among all the UEs. Numerical results show that, in scenarios with short scheduling blocks, the proposed framework outperforms the case where the DL and UL beamforming strategies are computed individually via two separate bi-directional training phases thanks to the reduced training overhead. Even more substantial gains are observed with respect to the case with a single bi-directional training phase, where the DL (resp. UL) beamforming strategies are reused in the UL (resp. DL).
\end{abstract}

\section{Introduction}

Massive multiple-input multiple-output (MIMO) is a multi-antenna technology at the foundation of the 5G New Radio (NR) standard. It enables highly directional beamforming and spatial multiplexing of many user equipments (UEs) thanks to massive antenna arrays at the base stations (BSs). However, the spectral efficiency of cell-edge UEs may still be compromised due to the high pathloss and interference from the neighboring BSs~\cite{Zha20}. In this context, cell-free massive MIMO aims to eliminate the inter-cell interference and ensure uniformly good service for all the UEs. This is realized by distributing a large number of antenna elements across the network and serving the UEs via coherent transmission/reception from all the BSs~\cite{Buz19,Int19,Raj20}. To this end, the BSs are assumed to be connected to a central processing unit (CPU) via backhaul links, which provides the UE-specific data and, possibly, the beamforming strategies adopted for the coherent transmission/reception.

Cell-free massive MIMO systems have been shown to outperform their traditional cellular massive MIMO and small-cell counterparts in several scenarios of practical interest~\cite{Ngo17,Bjo20}. In this context, it has been demonstrated that cooperative beamforming at the BSs can bring substantial gains with respect to local beamforming (such as simple matched filtering) at the expense of more intense signaling~\cite{Ngo17,Bjo20,Atz21,Gou20,Atz20}. In~\cite{Ngo17,Bjo20}, the cooperative beamforming strategies at the BSs are designed at the CPU for the case of single-antenna UEs. On the other hand, \cite{Atz21,Gou20,Atz20,Gou21} consider multi-antenna UEs and optimize the cooperative beamforming strategies at the BSs together with the beamforming vectors at the UEs in a distributed fashion building on bi-directional training~\cite{Tol19}. While a large part of the cell-free massive MIMO literature (including~\cite{Bjo20,Atz21,Gou20,Atz20}) focuses on the sum-rate performance, other works tackle the problem of maximizing the minimum signal-to-interference-plus-noise ratio (SINR) among all the UEs. In this regard, \cite{Ngo17} considers the max-min fairness power control problem for both downlink (DL) and uplink (UL) transmissions. Furthermore, \cite{Bas19} formulates the max-min SINR problem in the UL and optimizes the beamforming strategies at the BSs using geometric programming. A similar study is provided in \cite{Bas19a} assuming fronthaul capacity constraints. Lastly, \cite{Zho20} formulates the max-min SINR problem for the DL case and optimizes the beamforming strategies at the BSs exploiting the quasi-concavity of the minimum SINR objective.

The computation of the DL and UL beamforming strategies generally requires two separate bi-directional training phases~\cite{Tol19}. However, when the UEs to be served in the DL and in the UL are the same and the scheduling blocks are short, it may be convenient to devise a common beamforming strategy for both DL and UL that involves only one bi-directional training phase. In this paper, we tackle the joint optimization of the DL and UL beamforming strategies in cell-free massive MIMO systems with multi-antenna UEs. The proposed framework produces a set of power-constrained beamforming vectors that are used at the BSs and at the UEs for both DL and UL transmissions. In doing so, we consider the problem of maximizing the (weighted) minimum DL-UL rate among all the UEs. Our approach allows to avoid separate bi-directional training phases for the optimization of the DL and UL beamforming strategies, reduces the overall computational complexity, and can be adjusted to strike the desired balance between DL and UL performance. To facilitate the practical implementation of our method, we also present a heuristic approach that avoids additional signaling between the BSs and the UEs, which entails only a modest performance loss. Numerical results show that, in scenarios with short scheduling blocks, the proposed joint DL-UL design outperforms the case where the DL and UL beamforming strategies are computed individually via two separate bi-directional training phases. The observed gains are even higher with respect to the case where the DL (resp. UL) beamforming strategies are reused in the UL (resp. DL) direction.

\section{System Model and Problem Formulation}

Consider a cell-free massive MIMO network where a set of BSs $\setB \triangleq \{1, \ldots, B\}$, each equipped with $M$ antennas, serves a set of UEs $\setK \triangleq \{1, \ldots, K\}$, each equipped with $N$ antennas. The UEs are served in both the DL and the UL and receive/transmit up to $S$ data streams in each direction. Assuming a time division duplex (TDD) setting with channel reciprocity, let $\H_{b,k} \in \Compl^{M \times N}$ denote the UL channel matrix between UE~$k \in \setK$ and BS~$b \in \setB$, with $\H_{k} \triangleq [\H_{1,k}^{\tran}, \ldots, \H_{B,k}^{\tran}]^{\tran} \in \Compl^{B M \times N}$ representing the aggregate UL channel matrix of UE~$k$. Let $\w_{s,k} \in \Compl^{BM \times 1}$ be the aggregate beamforming vector used at all the BSs to transmit and receive stream~$s$ of UE~$k$. Similarly, let $\v_{s,k} \in \Compl^{N \times 1}$ be the beamforming vector used at UE~$k$ to transmit and receive stream~$s$. With the above definitions, the DL rate of UE~$k$ across all the streams can be written as
\begin{equation}\label{eq:R}
R^{\dl}_{k} \triangleq \sum_{s=1}^{S} \log_{2}(1 + \sinr^{\dl}_{s,k})
\end{equation}
where we have defined the DL SINR corresponding to stream~$s$ of UE~$k$ as
\begin{equation}\label{eq:SINR_k_dl}
\sinr^{\dl}_{s,k} \triangleq \frac{|\v_{s,k}^{\herm} \H_{k}^{\herm} \w_{s,k}|^{2}}{\sum_{(\bar{k},\bar{s}) \neq (k,s)} |\v_{s,k}^{\herm} \H_{k}^{\herm} \w_{\bar{s},\bar{k}}|^{2} + \sigma_{\ue}^{2} \| \v_{s,k} \|^{2}}
\end{equation}
where $\sigma_{\ue}^{2}$ is the additive white Gaussian noise (AWGN) variance at the UEs. Similarly, the UL rate of UE~$k$ across all the streams can be written as
\begin{equation}\label{eq:Rul}
R^{\ul}_{k} \triangleq \sum_{s=1}^{S} \log_{2}(1 + \sinr^{\ul}_{s,k})
\end{equation}
where we have defined the UL SINR corresponding to stream~$s$ of UE~$k$ as
\begin{align} \label{eq:SINR_k_ul}
\sinr^{\ul}_{s,k} & \triangleq \frac{| \w_{s,k}^{\herm} \H_{k} \v_{s,k}|^{2}}{\sum_{(\bar{k},\bar{s}) \neq (k,s)} | \w_{s,k}^{\herm} \H_{\bar{k}} \v_{\bar{s},\bar{k}}|^{2} + \sigma_{\bs}^{2} \|\w_{s,k}\|^{2}}
\end{align}
where $\sigma_{\bs}^{2}$ is the AWGN variance of the BS.

In this paper, we target the maximization of the minimum DL-UL rate using identical beamforming strategies for transmission and reception at the BSs and at the UEs.\footnote{Since scaling the combining vectors does not effect the SINR, any scaled version of the optimized beamforming vectors can be used at the receiver. Moreover, at the end of the beamforming training process, the receiver is free to use minimum mean squared error (MMSE) combining if required, which does not impact the beamforming training process.} The identical DL-UL beamforming design allows the joint optimization of the precoding-decoding strategies using a single bi-directional training mechanism and eliminates the need for separate bi-directional training phases for DL and UL. This reduces the overall training overhead and increases the effective data rate. The optimization problem that maximizes the minimum DL-UL rate with maximum transmit power at the BS and at the UEs $\rho_{\bs}$ and $\rho_{\ue}$, respectively, can be written~as

$ $ \vspace{-9.5mm}

\begin{align} \label{eq:prob_centr_1}
\begin{array}{cl}
\displaystyle \max_{\{\w_{s,k}, \v_{s,k}\}} & \displaystyle \min \big( \alpha \min_k R^{\dl}_{k}, (1-\alpha) \min_k R^{\ul}_{k} \big) \\
\mathrm{s.t.} & \displaystyle \sum_{k \in \setK} \sum_{s=1}^{S} \| \E_{b} \w_{s,k}\|^{2} \leq \rho_{\bs}, \quad \forall b \in \setB \\
& \displaystyle \sum_{s=1}^{S} \| \v_{s,k}\|^{2} \leq \rho_{\ue}, \quad \forall k \in \setK 
\end{array}
\end{align}
where $\E_{b} \in \Real^{M \times B M}$ is a selection matrix such that $\E_{b} \w_{s,k} = [\w_{s,k}((b-1)M+1),\ldots,\w_{s,k}(bM)]^{\tran}$ and $\alpha$ is the weight between DL and UL rates.

Let us introduce the auxiliary variables $\gamma_{s,k} \triangleq \sinr^{\dl}_{s,k}$ and $\bar \gamma_{s,k} \triangleq \sinr^{\ul}_{s,k}$. Hence, \eqref{eq:prob_centr_1} can be rewritten as \vspace{-1mm}
\begin{subequations} \label{eq:prob_centr_2}
\begin{align}
\displaystyle \max_{\substack{\{\w_{s,k}, \v_{s,k}\}, \\ \{\gamma_{s,k}, \bar \gamma_{s,k}\}, R}} & \ R \\
\label{eq:cen_rdl} \mathrm{s.t.} \hspace{5.5mm} & \ \alpha \sum_{s=1}^{S} \log_{2}(1 + \gamma_{s,k}) \ge R, \quad \forall k \in \setK \\
\label{eq:cen_ru} & \ (1-\alpha)\sum_{s=1}^{S} \log_{2}(1 + \bar \gamma_{s,k}) \ge R, \quad \forall k \in \setK \\
\label{eq:cen_sinr_dl} & \ \eqref{eq:SINR_k_dl} \ge \gamma_{s,k}, \quad \forall k \in \setK,~s = 1, \ldots, S \\
\label{eq:cen_sinr_ul} & \ \eqref{eq:SINR_k_ul} \ge \bar \gamma_{s,k}, \quad \forall k \in \setK,~s = 1, \ldots, S \\
\label{eq:cen_bs_pwr} & \ \displaystyle \sum_{k \in \setK} \sum_{s=1}^{S} \| \E_{b} \w_{s,k}\|^{2} \leq \rho_{\bs}, \quad \forall b \in \setB \\
\label{eq:cen_ue_pwr} & \ \sum_{s=1}^{S} \| \v_{k}\|^{2} \leq \rho_{\ue}, \quad \forall k \in \setK
\end{align}
\end{subequations}
where $R$ is the minimum common rate in~the~DL~and~in~the~UL. We apply alternating optimization to~\eqref{eq:prob_centr_2} to optimize \{$\w_{s,k}\}$ and $\{\v_{s,k}\}$. Specifically, we derive the optimal BS beamformers $\{\w_{s,k}\}$ for fixed $\{\v_{s,k}\}$ in Section~\ref{sec:BS_bf} and the optimal UE beamformers $\{\v_{s,k}\}$ for fixed $\{\w_{s,k}\}$ in Section~\ref{sec:UE_bf}.

\vspace{-0.5mm}

\section{BS Beamforming Design}\label{sec:BS_bf}

\vspace{-0.5mm}

For a fixed set of UE beamformers $\{\v_{s,k}\}$, the BS beamformers $\{\w_{s,k}\}$ are obtained by solving \vspace{-1.5mm}
\begin{align}\label{eq:prob_bs}
\begin{array}{cl}
\displaystyle \max_{\{\w_{s,k}, \gamma_{s,k}, \bar \gamma_{s,k}\}, R} & R \\ 
\mathrm{s.t.} & \textnormal{\eqref{eq:cen_rdl}--\eqref{eq:cen_bs_pwr}}.
\end{array}
\end{align}
The constraints in~\eqref{eq:cen_sinr_dl} and~\eqref{eq:cen_sinr_ul} are non-convex with respect to $\{\w_{s,k}\}$, $\{\gamma_{s,k}\}$, and $\{\bar \gamma_{s,k}\}$. Hence, we use successive convex approximation (SCA)~\cite{Scu17} based on the first-order approximation of~\eqref{eq:cen_sinr_dl} and~\eqref{eq:cen_sinr_ul}.

We begin by introducing the auxiliary variables $p_{s,k}(\w_{s,k},\gamma_{s,k}) \triangleq \frac{1}{\gamma_{s,k}}|\v_{s,k}^{\herm} \H_{k}^{\herm}\w_{s,k}|^{2}$ and $q_{s,k}(\w_{s,k},\bar \gamma_{s,k}) \triangleq \frac{1}{\bar \gamma_{s,k}}| \w_{s,k}^{\herm} \H_{k} \v_{s,k}|^{2}$. The first-order approximations of $p_{s,k}$ and $q_{s,k}$ at iteration $(i+1)$ are given, respectively, by \vspace{-1mm}
\begin{align}
\tilde p_{s,k}(\w_{s,k},\gamma_{s,k}) & \triangleq - \frac{|\v_{s,k}^{\herm} \H_{k}^{\herm} \w^{(i)}_{s,k}|^{2}}{\gamma_{s,k}^{(i)}} \frac{\gamma_{s,k} }{\gamma_{s,k}^{(i)}} \nonumber\\
& \phantom{=} \ + 2 \frac{{(\w_{s,k}^{(i)})^{\herm}} \H_{k}\v_{s,k}\v_{s,k}^{\herm} \H_{k}^{\herm} \w_{s,k}}{\gamma_{s,k}^{(i)}}
\end{align}

\vspace{-2mm}

\noindent and

$ $ \vspace{-9mm}

\begin{align}
\tilde q_{s,k}(\w_{s,k},\bar \gamma_{s,k}) & = - \frac{\big|(\w_{s,k}^{(i)})^{\herm} \H_{ k} \v_{s, k} \big|^{2}}{\bar \gamma_{s,k}^{(i)}} \frac{\bar \gamma_{s,k} }{\bar \gamma_{s,k}^{(i)}} \nonumber\\
& \phantom{=} \ + 2 \frac{{(\w_{s,k}^{(i)})^{\herm}} \H_{ k}\v_{s,k}\v_{s,k}^{\herm} \H_{k}^{\herm} \w_{s,k}}{\bar \gamma_{s,k}^{(i)}}.
\end{align}
Since $p_{s,k}$ and $q_{s,k}$ are convex, it follows that $\tilde p_{s,k} \le p_{s,k}$ and $\tilde q_{s,k} \le q_{s,k}$. Therefore, we approximate~\eqref{eq:cen_sinr_dl} and~\eqref{eq:cen_sinr_ul} as
\vspace{-0.5mm}
\begin{align} 
\label{eq:dlapprx_bs}{\sum_{(\bar{k},\bar{s}) \neq (k,s)} |\v_{s,k}^{\herm} \H_{k}^{\herm} \w_{\bar{s},\bar{k}}|^{2} + \sigma_{\ue}^{2} \| \v_{s,k} \|^{2}} & \le \tilde p_{s,k}, \\
\label{eq:ulapprx_bs}{\sum_{(\bar{k},\bar{s}) \neq (k,s)} | \w_{s,k}^{\herm} \H_{\bar{k}} \v_{\bar{s},\bar{k}}|^{2} + \sigma_{\bs}^{2} \|\w_{s,k}\|^{2}} & \le \tilde q_{s,k}
\end{align}
respectively, which yields the following convex approximation of~\eqref{eq:prob_bs}: \vspace{-2mm}
\begin{subequations}\label{eq:prob_bs_appx}
\begin{align}
\displaystyle \max_{\{\w_{s,k}, \gamma_{s,k}, \bar \gamma_{s,k}\}, R} & \ R \\ 
\mathrm{s.t.} \hspace{9mm} & \ \textnormal{\eqref{eq:cen_rdl}--\eqref{eq:cen_ru}~and~\eqref{eq:cen_bs_pwr}} \\
\label{eq:bs_dlsinr_cvx_apx} & \ \eqref{eq:dlapprx_bs}, \quad \forall k \in \setK,~s = 1, \ldots, S \\
\label{eq:bs_ulsinr_cvx_apx} & \ \eqref{eq:ulapprx_bs}, \quad \forall k \in \setK,~s = 1, \ldots, S.
\end{align}
\end{subequations}
The optimal BS beamformer $\w_{s,k}$ can be obtained by finding the stationary point of the Lagrangian of~\eqref{eq:prob_bs_appx}, and the optimal $\w_{s,k}$ is given by \vspace{-1mm}
\begin{align}\label{eq:ideal_bs_bf}
\w_{s,k} & \triangleq \bigg(\sum_{(\bar{k},\bar{s}) \neq (k,s)} \big(\alpha \nu_{\bar{s}, \bar{k}} + (1-\alpha) \mu_{s,k} \big) \H_{\bar{k}} \v_{\bar{s}, \bar{k}} \v_{\bar{s}, \bar{k}}^{\herm} \H_{\bar{k}}^{\herm} \nonumber \\
& \phantom{=} \ + (1-\alpha) \mu_{s,k} \sigma_{\bs}^{2} \I_{BM} + \sum_{b \in \setB}\lambda_b \E^{\herm}_{b}\E_{b} \bigg)^{-1} \nonumber \\
& \phantom{=} \ \times \bigg( \bigg( \alpha \frac{\nu_{s,k}}{\gamma_{s,k}^{(i)}} + (1-\alpha) \frac{ \mu_{s,k}}{{\bar \gamma_{s,k}^{(i)}}} \bigg) \H_{k}\v_{s, k}\v_{s,k}^{\herm} \H_{k}^{\herm} \w_{s,k}^{(i)} \bigg)
\end{align}
where $\nu_{s,k}$, $\mu_{s,k}$, and $\lambda_b$ are the dual variables corresponding to~\eqref{eq:bs_dlsinr_cvx_apx}, \eqref{eq:bs_ulsinr_cvx_apx}, and~\eqref{eq:cen_bs_pwr}, respectively. As detailed in the Appendix, $\mu_{s,k}$ and $\nu_{s,k}$ depend on the auxiliary dual variables $\eta_k$ and $\zeta_k$ corresponding to~\eqref{eq:cen_rdl} and~\eqref{eq:cen_ru}, respectively, and are updated using the sub-gradient method.

\subsection{BS Beamforming Design with UL Training} \label{sec:BS_opt}

Let $\h_{s,k} \triangleq \H_{k} \v_{s,k} \in \Compl^{BM \times 1}$ denote the effective UL channel vector between stream~$s$ of UE~$k$ and all the BSs, and let $\p_{s,k} \in \Compl^{\tau \times 1}$ denote the pilot assigned to stream~$s$ of UE~$k$, with $\|\p_{s,k}\|^{2} = \tau$. During the UL pilot-aided channel estimation phase, each UE~$k$ synchronously transmits a superposition of its pilots $\{\p_{s,k}\}_{s=1}^{S}$ precoded with the corresponding $\{\v_{s,k}\}_{s=1}^{S}$, i.e., \vspace{-1mm}
\begin{align} \label{eq:X_k_ul1}
\X_{k}^{\ul} \triangleq \sum_{s=1}^{S}\v_{s,k} \p_{s,k}^{\herm} \in \Compl^{N \times \tau}.
\end{align}
Then, the receive signal across all the BSs is given by \vspace{-0.5mm}
\begin{align}
\Y^{\ul} & \triangleq \sum_{k \in \setK} \H_{k} \X_{k}^{\ul} + \Z^{\ulA} \\
 & = \sum_{k \in \setK} \sum_{s=1}^{S} \h_{s,k} \p_{s,k}^{\herm} + \Z^{\ulA} \in \Compl^{BM \times \tau}
\end{align}
\noindent where $\Z^{\ulA}$ is the AWGN term across all the BSs, and the least-squares (LS) estimate of~$\h_{s,k}$ is obtained as
\begin{align}
\hat{\h}_{s,k} & \triangleq \frac{1}{\tau} \Y^{\ul} \p_{s,k} \\
\label{eq:h_hat} & = \h_{s,k} + \frac{1}{\tau} {\sum_{(\bar{k},\bar{s}) \neq (k,s)}} \h_{\bar s,\bar{k}} \p_{\bar s, \bar{k}}^{\herm} \p_{s,k} + \frac{1}{\tau} \Z^{\ulA} \p_{s,k}.
\end{align}

With the estimated channel $\hat{\h}_{s,k}$, the BS beamformer $\w_{s,k}$ can be computed at the CPU as
\begin{align}\label{eq:esti_bs_bf}
\hat \w_{s,k} & \triangleq \bigg(\sum_{(\bar{k},\bar{s}) \neq (k,s)} \big(\alpha \nu_{\bar{s}, \bar{k}} + (1-\alpha) \mu_{s,k} \big) \hat \h_{\bar s, \bar{k}} \hat \h_{\bar s, \bar{k}}^{\herm} \nonumber \\
& \phantom{=} \ + (1-\alpha) \mu_{s,k} \sigma_{\bs}^{2} \I_{BM} + \sum_{b \in \setB}\lambda_b \E^{\herm}_{b}\E_{b} \bigg)^{-1} \nonumber \\
& \phantom{=} \ \times \bigg( \bigg( \alpha \frac{\nu_{s,k}}{\gamma_{s,k}^{(i)}} + (1-\alpha) \frac{ \mu_{s,k}}{{\bar \gamma_{s,k}^{(i)}}} \bigg) \hat \h_{s,k}\hat \h_{s,k}^{\herm} \w_{s,k}^{(i)} \bigg).
\end{align}
Note that~\eqref{eq:esti_bs_bf} becomes equal to~\eqref{eq:ideal_bs_bf} as $\tau \to \infty$. The dual variables~$\nu_{s,k}$ and~$\mu_{s,k}$ are computed at the CPU using the estimated effective channel in \eqref{eq:h_hat}, i.e., by replacing $\H_k\v_{s,k}$ with $\hat \h_{s,k}$ in \eqref{eq:nu}--\eqref{eq:mu} (see the Appendix). Similarly, $\gamma_{s,k}$ and $\bar \gamma_{s,k}$ are obtained by plugging the estimated effective channels into \eqref{eq:SINR_k_dl} and \eqref{eq:SINR_k_ul}, respectively. For each bi-directional training iteration, $\hat \w_{s,k}$ is computed via the sub-gradient method until the convergence of the auxiliary dual variables $\eta_k$ and $\zeta_k$.

\begin{figure*}[t!]
\vspace{-1mm}

\addtocounter{equation}{+13}
\begin{align}\label{eq:esti_ue_bf}
\nonumber \hat \v_{s,k} & \triangleq \bigg( \alpha \bar{\nu}_{s,k} \big( \Y_{k}^{\dlA} (\Y_{k}^{\dlA})^{\herm} \! - \! \Y_{k}^{\dlA}\p_{s,k} \p_{s,k}^{\herm} (\Y_{k}^{\dlA})^{\herm} \! - \! \tau \sigma^2_{\ue}\I_{N} \big) \! + \! (1 \! - \! \alpha) \big( \Y_{k}^{\dlB} (\Y_{k}^{\dlB})^{\herm} \! - \! \Y_{k}^{\dlB}\p_{s,k} \p_{s,k}^{\herm} (\Y_{k}^{\dlB})^{\herm} \! - \! \tau \sigma^2_{\ue}\I_{N} \big) \\
& \phantom{=} \ + \! \alpha\bar{\nu}_{s,k} \sigma_{\ue}^{2}\tau \I_{N} \! + \! \bar{\lambda}_k \I_{N} \bigg)^{-1} \bigg( \alpha \frac{\bar{\nu}_{s,k}}{\gamma_{s,k}^{(i)}} \Y_{k}^{\dlA}\p_{s,k} \p_{s,k}^{\herm} (\Y_{k}^{\dlA})^{\herm} {\v^{(i)}_{s,k}} \! + \! (1 \! - \! \alpha) \frac{1}{\bar \gamma_{s,k}^{(i)}} \Y_{k}^{\dlB}\p_{s,k} \p_{s,k}^{\herm} (\Y_{k}^{\dlB})^{\herm} {\v^{(i)}_{s,k}} \bigg)
\end{align}
\addtocounter{equation}{-14}

\vspace{-5mm}

\hrulefill

\vspace{-4mm}
\end{figure*}

\section{UE Beamforming Design}\label{sec:UE_bf}
 
For a fixed set of BS beamformers $\{\w_{s,k}\}$, the UE beamformers $\{\v_{s,k}\}$ are obtained by solving
\begin{align}\label{eq:prob_ue}
\begin{array}{cl}
\displaystyle \max_{\{\v_{s,k}, \gamma_{s,k}, \bar \gamma_{s,k}\}, R} & R \\ 
\mathrm{s.t.} & \textnormal{\eqref{eq:cen_rdl}--\eqref{eq:cen_sinr_ul}~and~\eqref{eq:cen_ue_pwr}}.
\end{array}
\end{align}
The constraints in~\eqref{eq:cen_sinr_dl} and~\eqref{eq:cen_sinr_ul} are non-convex with respect to $\{\v_{s,k}\}$, $\{\gamma_{s,k}\}$, and $\{\bar \gamma_{s,k}\}$. Therefore, as done in Section~\ref{sec:BS_bf}, we resort to SCA.

We begin by introducing the auxiliary variables $r_{s,k}(\v_{s,k},\gamma_{s,k}) \triangleq \frac{|\v_{s,k}^{\herm} \H_{k}^{\herm} \w_{s,k}|^{2}}{\gamma_{s,k}}$ and $t_{s,k}(\v_{s,k},\bar \gamma_{s,k}) \triangleq \frac{| \w_{s,k}^{\herm} \H_{k} \v_{s,k}|^{2}}{\bar \gamma_{s,k}}$. The first-order approximations of $r_{s,k}$ and $t_{s,k}$ at iteration $(i+1)$ are given, respectively, by
\begin{align}
\tilde{r}_{s,k}(\v_{s,k},\gamma_{s,k}) & \triangleq - \frac{\big|(\v_{s,k}^{(i)})^{\herm} \H_{k}^{\herm} \w_{s, k}\big|^{2}}{{\gamma_{s,k}}^{(i)}} \frac{\gamma_{s,k} }{\gamma_{s,k}^{(i)}} \nonumber \\
& \phantom{=} \ + 2 \frac{{(\v_{s,k}^{(i)})^{\herm}} \H_{k}^{\herm}\w_{s,k}\w_{s,k}^{\herm} \H_{k} \v_{s,k}}{\gamma_{s,k}^{(i)}},\\
\tilde{t}_{s,k}(\v_{s,k},\bar \gamma_{s,k}) & \triangleq - \frac{|{\w_{s,k}^{\herm}} \H_{ k} \v^{(i)}_{s, k}|^{2}}{\bar \gamma_{s,k}^{(i)}} \frac{\bar \gamma_{s,k} }{{\bar \gamma_{s,k}^{(i)}}} \nonumber \\
& \phantom{=} \ + 2 \frac{{(\v_{s,k}^{(i)})^{\herm}} \H_{k}^{\herm} \w_{s,k} \w_{s,k}^{\herm} \H_{k} \v_{s,k}}{{\bar \gamma_{s,k}^{(i)}}}.
\end{align}
Since $r_{s,k}$ and $t_{s,k}$ are convex, it follows that $\tilde r_{s,k} \le r_{s,k}$ and $\tilde t_{s,k} \le t_{s,k}$. Therefore, we approximate~\eqref{eq:cen_sinr_dl} and~\eqref{eq:cen_sinr_ul} as

$ $ \vspace{-10mm}

\begin{align} 
\label{eq:dlapprx}{\sum_{(\bar{k},\bar{s}) \neq (k,s)} |\v_{s,k}^{\herm} \H_{k}^{\herm} \w_{\bar{s},\bar{k}}|^{2} + \sigma_{\ue}^{2} \| \v_{s,k} \|^{2}} & \le \tilde r_{s,k}, \\
\label{eq:ulapprx}{\sum_{(\bar{k},\bar{s}) \neq (k,s)} | \w_{s,k}^{\herm} \H_{\bar{k}} \v_{\bar{s},\bar{k}}|^{2} + \sigma_{\bs}^{2} \|\w_{s,k}\|^{2}} & \le \tilde t_{s,k}
\end{align}
respectively, which yields the following convex approximation of~\eqref{eq:prob_ue}:
\begin{subequations}\label{eq:prob_ue_appx}
\begin{align}
\displaystyle \max_{\{\v_{s,k}, \gamma_{s,k}, \bar \gamma_{s,k}\}, R} & \ R \\
\mathrm{s.t.} \hspace{9mm} & \ \textnormal{\eqref{eq:cen_rdl}--\eqref{eq:cen_ru}~and~\eqref{eq:cen_ue_pwr}} \\
\label{eq:dlsinr_cvx_apx} & \ \eqref{eq:dlapprx}, \quad \forall k \in \setK,~s = 1, \ldots, S \\
\label{eq:ulsinr_cvx_apx} & \ \eqref{eq:ulapprx}, \quad \forall k \in \setK,~s = 1, \ldots, S.
\end{align}
\end{subequations}
The optimal UE beamformer $\v_{s,k}$ can be obtained by finding the stationary point of the Lagrangian of~\eqref{eq:prob_ue_appx}, and the optimal $\v_{s,k}$ is given by
\begin{align}\label{eq:ideal_ue_bf}
\v_{s,k} & \triangleq \bigg( \sum_{(\bar{k},\bar{s}) \neq (k,s)} \big( \alpha\bar{\nu}_{s,k} +(1-\alpha) \bar{\mu}_{\bar{s}, \bar{k}} \big) \H_{{k}}^{\herm} \w_{\bar{s}, \bar{k}} \w_{\bar{s},\bar{k}}^{\herm} \H_{k} \nonumber \\
& \phantom{=} \ + \alpha\bar{\nu}_{s,k} \sigma_{\ue}^{2}\I_N + \bar{\lambda}_k \I_N \bigg)^{-1} \bigg( \bigg( \alpha\frac{\bar{\nu}_{s,k}}{\gamma_{s,k}^{(i)}} + (1-\alpha) \frac{\bar{\mu}_{s,k}}{\bar \gamma_{s,k}^{(i)}} \bigg) \nonumber \\
& \phantom{=} \ \times \H_{k}^{\herm} \w_{s,k} \w_{s,k}^{\herm} \H_{k} {\v_{s,k}}^{(i)} \bigg)
\end{align}
where $\bar{\nu}_{s,k}$, $\bar{\mu}_{s,k}$, and $\bar{\lambda}_k$ are the dual variables corresponding to~\eqref{eq:dlsinr_cvx_apx}, \eqref{eq:ulsinr_cvx_apx}, and~\eqref{eq:cen_ue_pwr}, respectively. As detailed in the Appendix, $\bar{\mu}_{s,k}$ and $\bar{\nu}_{s,k}$ depend on the auxiliary dual variables $\eta_k$ and $\zeta_k$ corresponding to~\eqref{eq:cen_rdl} and~\eqref{eq:cen_ru}, respectively. Following similar steps as in Section~\ref{sec:BS_opt}, the latter can be updated using the sub-gradient method. However, this requires additional DL signaling, as described next in Section~\ref{sec:UE_opt}. Alternatively, one can use the heuristic approach proposed in Section~\ref{sec:UE_heur} to avoid both the sub-gradient update (which may be computationally complex at the UEs) and the~additional~DL~signaling.

\subsection{UE Beamforming Design with DL Training} \label{sec:UE_opt}

During the DL pilot-aided channel estimation phase, each BS~$b$ synchronously transmits a superposition of the pilots $\{\p_{s,k}\}$ precoded with the corresponding $\{\w_{s,k}\}$, i.e.,

$ $ \vspace{-9mm}

\begin{align} \label{eq:X_b_dlA}
\X^{\dlA} \triangleq \sum_{k \in \setK} \sum_{s=1}^{S} \w_{s,k} \p_{s,k}^{\herm} \in \Compl^{BM \times \tau}.
\end{align}
Then, the receive signal at UE~$k$ is given by
\begin{align}\label{eq:y_b_dlA}
\Y_{k}^{\dlA} & \triangleq \H_{k}^{\herm} \X^{\dlA} + \Z_{k}^{\dlA} \\
\label{eq:Y_k_dlA} & = \sum_{\bar k \in \setK} \sum_{s=1}^{S} \H_{k}^{\herm} \w_{s,\bar{k}} \p_{s,\bar{k}}^{\herm} + \Z_{k}^{\dlA} \in \Compl^{N \times \tau}
\end{align}
where $\Z_{k}^{\dlA}$ is the AWGN term at UE~$k$. To allow each UE to independently compute its beamforming strategy, the BS needs to transmit a superposition of precoded pilots weighted with the corresponding dual variables $\{\bar \mu_{s,k}\}$, i.e.,
\begin{align} \label{eq:X_b_dlB}
\X^{\dlB} \triangleq \sum_{k \in \setK} \sum_{s=1}^{S} \sqrt{\bar \mu_{s,k}}\w_{s,k} \p_{s,k}^{\herm} \in \Compl^{BM \times \tau}
\end{align}
and the receive signal at UE~$k$ is given by
\begin{align}
\Y_{k}^{\dlB} & \triangleq \H_{k}^{\herm} \X^{\dlB} + \Z_{k}^{\dlB} \\
\label{eq:Y_k_dlB} & = \sum_{\bar k \in \setK} \sum_{s=1}^{S} \sqrt{\bar \mu_{s,\bar k}} \H_{k}^{\herm} \w_{s,\bar{k}} \p_{s,\bar{k}}^{\herm} + \Z_{k}^{\dlB} \in \Compl^{N \times \tau}
\end{align}
where $\Z_{k}^{\dlB}$ is the AWGN term at UE~$k$. Here, the dual variable~$\bar \mu_{s,k}$ is computed at the CPU as described in Section~\ref{sec:BS_opt}.

Building on~\eqref{eq:Y_k_dlA} and~\eqref{eq:Y_k_dlB}, the UE beamformer $\v_{s,k}$ can be computed as in \eqref{eq:esti_ue_bf} at the top of the page. Note that~\eqref{eq:esti_ue_bf} becomes equal to \eqref{eq:ideal_ue_bf} as $\tau \to \infty$. The dual variable $\bar \nu_{s,k}$ is computed at UE~$k$ by~replacing $\H_k\w_{s,k}$ with $\frac{1}{\tau} \Y_{k}^{\dlA}\p_{s,k}$~in \eqref{eq:nu_bar} (see the Appendix). For $\alpha=1$ and no transmit~power constraint at UE~$k$ (i.e., $\bar{\lambda}_{k}=0$), $\hat \v_{s,k}$ in~\eqref{eq:esti_ue_bf} is equivalent to the MMSE receiver, which can be implemented~based~solely~on $\Y_{k}^{\dlA}$ (i.e., without $\Y_{k}^{\dlB}$). Similarly, for $\alpha=0$, $\hat \v_{s,k}$ in~\eqref{eq:esti_ue_bf} is equivalent to UL max-min rate transmit beamforming, which can implemented based solely on $\Y_{k}^{\dlB}$ (i.e., without $\Y_{k}^{\dlA}$).

\subsection{Heuristic UE Beamforming Design} \label{sec:UE_heur}

Let us consider the ideal UE beamforming design given in~\eqref{eq:ideal_ue_bf} and let us define $a_{\bar s,\bar k}(s,k) \triangleq {\alpha \bar\nu_{s,k} + (1-\alpha)\bar\mu_{\bar s,\bar k}} $ and $b_{s,k} \triangleq {\alpha \bar \nu_{s,k}}$. Building on these definitions, \eqref{eq:ideal_ue_bf} can be equivalently written as
\begin{align}\label{eq:hrst_ue_bf}
\addtocounter{equation}{+1}
\v_{s,k} & = \bigg( \sum_{(\bar{k},\bar{s})} a_{\bar{s},\bar{k}}(s,k) \H_{{k}}^{\herm} \w_{\bar{s}, \bar{k}} \w_{\bar{s},\bar{k}}^{\herm} \H_{k} \nonumber \\
& \phantom{=} \ + (b_{s,k} \sigma_{\ue}^{2} + \bar{\lambda}_k) \I_{N} \bigg)^{-1} \H_{k} \w_{s,k}.
\end{align}

As in Section~\ref{sec:UE_opt}, the practical implementation of~\eqref{eq:hrst_ue_bf} requires additional DL signaling, i.e., $\Y_{k}^{\dlA}$ and $\Y_{k}^{\dlB}$. Furthermore, the sub-gradient update of the auxiliary dual variables slows down the overall convergence. First, to avoid the additional DL signaling, we consider a heuristic approach for the UE beamformer design whereby the values of $a_{\bar s, \bar k}(s,k)$ are assumed to be the same for all the UE beamformers and streams, i.e., $a_{\bar{s}, \bar{k}}(s, k) = a_{\bar{s}, \bar{k}},$ $\forall s, k$. In this setting, the heuristic UE beamforming design can be implemented based on $\Y_{k}^{\dlB}$ (obtained by replacing $\bar \mu_{s,k}$~with~$a_{s,k}$~in~\eqref{eq:X_b_dlB})~as
\begin{align}\label{eq:esti_ue_bf_prac}
\hat \v_{s,k} \! & = \! \big( \Y_{k}^{\dlB} (\Y_{k}^{\dlB})^{\herm} \!- \! \tau(1 \! - \! b_{s,k})\sigma^2_{\ue}\I_{N} \! + \! \bar{\lambda}_k \I_{N} \big)^{-1} \frac{\Y_{k}^{\dlB}\p_{s,k}}{\sqrt{a_{s,k}}}.
\end{align}
Second, to avoid the sub-gradient update of the auxiliary dual variables at the UEs, $a_{s,k}$ and $b_{s,k}$ are heuristically updated at the BSs (CPU) and the UEs. This leads to a faster convergence of the heuristic method to a sub-optimal point. As an example, a simple heuristic UE beamformer can be obtained by setting $\{a_{s,k}\!=\!1\}$ and $\{b_{s,k}\!=\!0\}$. Otherwise, one can tune $\{a_{s,k}\}$ and $\{b_{s,k}\}$ for faster and better convergence based on the statistics of the network parameters such as the number of the BSs/UEs, the transmit power at the BSs/UEs, and the channel~conditions.

\section{Numerical Results}

We consider a cell-free massive MIMO scenario where $B=25$~BSs, each equipped with $M=4$~antennas, are placed on a square grid with distance between neighboring BSs of $100$~m. Furthermore, $K=16$~UEs, each equipped with $N = 2$~antennas and served with $S=2$ data streams, are randomly dropped in the same area. Assuming uncorrelated Rayleigh fading, each channel is generated according to $\mathrm{vec}(\H_{b,k}) \sim \setC \setN (0, \delta_{b,k} \I_{MN})$, where $\delta_{b,k} \triangleq -61.3-30 \log_{10} (d_{b,k}) -20 \log_{10} \big( f_c~\textrm{[GHz]} \big)$ [dB] is the large-scale fading coefficient, $d_{b,k}$ is the distance between BS~$b$ and UE~$k$, and $f_c=28$~GHz is the carrier frequency~\cite{Lee15}. The maximum transmit power for both the data and the pilot transmission is $\rho_{\bs} = 30$~dBm at the BSs and $\rho_{\ue} = 20$~dBm at the UEs. The AWGN power at the BSs and at the UEs is fixed to $\sigma_{\bs}^{2} = \sigma_{\ue}^{2} = -95$~dBm. Lastly, we set $\alpha = 0.5$ for equal weight between DL and UL rates (see \eqref{eq:prob_centr_1}).

We begin by considering the ideal scenario where the CPU and each UE have global knowledge of the channels and of the previously computed beamforming strategies. The performance of the proposed joint DL-UL design, termed as \textit{DL-UL opt.}, is compared with the case where the DL and UL beamforming strategies are computed individually via two separate bi-directional training phases, termed as \textit{Separate DL/UL opt.} We also provide a comparison with the cases where the DL (resp. UL) beamforming strategies are computed via a single bi-directional training phase and are reused in the UL (resp. DL) direction, termed as \textit{DL opt.} (resp. \textit{UL opt.}). The practical versions of these schemes that avoid the sub-gradient updates at the UEs and the additional DL signaling are referred to as \textit{DL-UL heur.}, \textit{Separate DL/UL heur.}, and \textit{UL heur.}, respectively (see Section~\ref{sec:UE_heur}).

\begin{figure}[t!]
\vspace{-3mm}
\centering
\includegraphics[width=0.9\columnwidth]{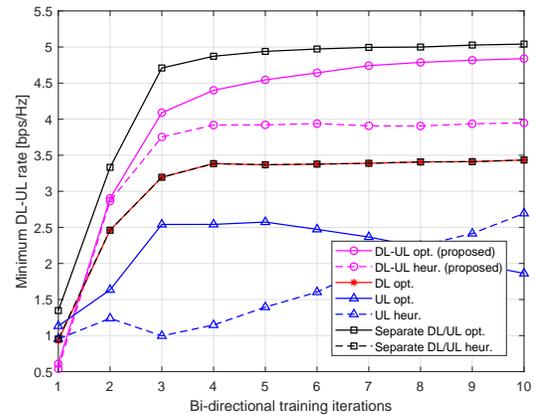}
\vspace{-3mm}
\caption{Minimum DL-UL rate versus the number of bi-directional training iterations.} \vspace{-5mm}
\label{fig:minrate}
\end{figure}

As shown in Figure~\ref{fig:minrate}, the minimum DL-UL rate of the proposed \textit{DL-UL opt.} approaches that of the \textit{Separate DL/UL opt.} However, the latter requires two separate bi-directional training phases for the DL and UL beamforming design. Furthermore, the \textit{DL-UL opt.} outperforms the \textit{DL opt.} and the \textit{UL opt.}, where all these methods are characterized by the same training requirements. A similar trend can be also observed for the practical heuristic UE beamforming design. In this respect, the proposed \textit{DL-UL heur.} outperforms the \textit{DL opt.}, the \textit{UL heur.}, and the \textit{Separate DL/UL heur.}

\begin{figure}[t!]
\centering
\includegraphics[width=0.9\columnwidth]{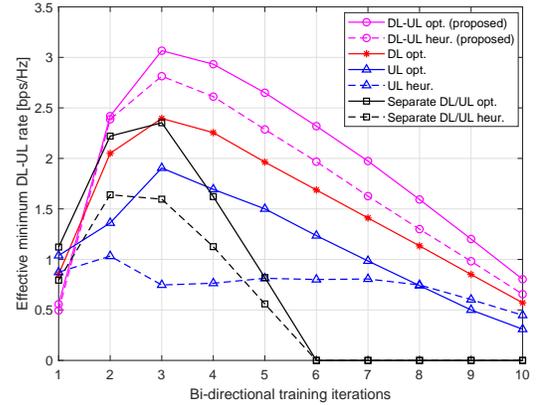}
\vspace{-3mm}
\caption{Effective minimum DL-UL rate versus the number of bi-directional training iterations with scheduling block size of 4 slots.} \vspace{-5mm}
\label{fig:effrate}
\end{figure}

Figure~\ref{fig:effrate} plots the effective minimum DL-UL rate taking into account the bi-directional training overhead with scheduling block size of 4 slots. While the rate improves with the number of bi-directional training iterations, the fraction of time left for the actual data transmission is reduced, giving rise to a clear performance trade-off. At the optimal number of iterations, the effective minimum DL-UL rate of the proposed \textit{DL-UL opt.} is 25\% higher than that of the \textit{Separate DL/UL opt.}, since the latter involves two separate bi-directional training phases for the DL and UL beamforming design. As expected, the effective minimum DL-UL rates produced by the \textit{DL opt.} and the \textit{UL opt.} are lower than that of the \textit{DL-UL opt.}, where all these methods are characterized by the same training requirements. This behavior can be also seen in the practical heuristic UE beamforming design, where the proposed \textit{DL-UL heur.} improves the effective minimum DL-UL rate by 17\% with respect to the \textit{DL opt.} More significant gains are observed when comparing to the \textit{UL heur.} and the \textit{Separate DL/UL~heur.}


\begin{figure}[t!]
\vspace{-3mm}
\centering
\includegraphics[width=0.9\columnwidth]{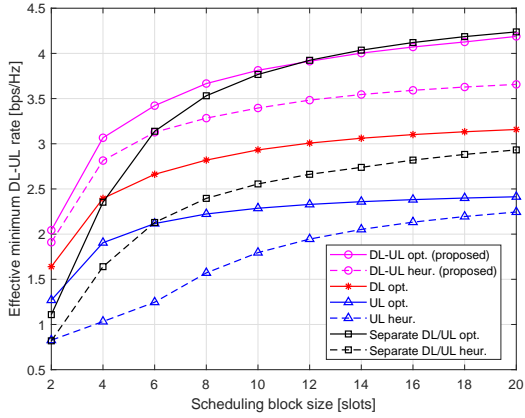}
\vspace{-3mm}
\caption{Effective minimum DL-UL rate versus the scheduling block size.} \vspace{-5mm}
\label{fig:EffratevsSB}
\end{figure}

In Figure~\ref{fig:EffratevsSB}, the optimal effective minimum DL-UL rate is plotted as a function of the scheduling block size. For short scheduling blocks, the proposed \textit{DL-UL opt.} outperforms the \textit{Separate DL/UL opt.}, the \textit{DL opt.}, and the \textit{UL opt.} However, for large scheduling blocks (i.e., larger than 12 slots), the \textit{Separate DL/UL opt.} is superior to our joint DL-UL design, since the impact of the bi-directional training reduces with the increasing scheduling block size. Even in the practical heuristic UE beamforming design, the proposed \textit{DL-UL heur.} surpasses the \textit{DL opt.}, the \textit{UL heur.}, and the \textit{Separate DL/UL heur.}, as well as the \textit{Separate DL/UL opt.} for scheduling blocks shorter than 6 slots. Moreover, the effective minimum DL-UL rate of the \textit{DL opt.} is better than that of the \textit{Separate DL/UL heur.} In fact, the two schemes have the same minimum DL-UL rate (as seen in Figure~\ref{fig:minrate}) but the latter requires more training.

\section{Conclusions}
The computation of the DL and UL beamforming strategies in cell-free massive MIMO systems generally requires~two separate~bi-directional training phases.Considering the scenario where the UEs to be served in the DL and in the UL are the same and the scheduling blocks are short, we jointly optimize the DL and UL beamforming strategies using only one bi-directional training phase. The proposed framework maximizes the (weighted) minimum DL-UL rate among all the UEs and produces a set of power-constrained beamforming vectors that are used at the BSs and at the UEs for both DL and UL transmissions. Moreover, we present a heuristic approach that avoids additional signaling between the BSs and the UEs, which entails only a modest performance loss.~For~short scheduling blocks, the reduced training overhead of the proposed joint DL-UL optimization brings substantial gains (up to 25\%) with respect to the case where the DL and UL beamforming strategies are computed individually via two separate bi-directional training phases. Even higher gains are observed with respect to the case where the DL~(resp. UL) beamforming strategies are reused in the UL (resp. DL)~direction.

\section*{Appendix}\label{sec:appendix}

The dual variables corresponding to~\eqref{eq:cen_rdl} and~\eqref{eq:cen_ru} are given by \vspace{-1.5mm}
\begin{align}
\eta_k & \triangleq \eta_{k}^{(i)} - \delta (\alpha \sum_{s=1}^{S} \log_{2}(1 + \gamma_{s,k}^{(i)}) - R), \\
\zeta_k & \triangleq \zeta_{k}^{(i)} - \delta ((1-\alpha) \sum_{s=1}^{S} \log_{2}(1 + \bar \gamma_{s,k}^{(i)}) - R)
\end{align}
respectively, where $\delta$ is the sub-gradient step size. In addition, the dual variables corresponding to~\eqref{eq:bs_dlsinr_cvx_apx} and~\eqref{eq:bs_ulsinr_cvx_apx} are defined as \vspace{-1.5mm}
\begin{align}
\label{eq:nu} \nu_{s,k} & \triangleq \frac{ \eta_k\alpha (\gamma_{s,k}^{(i)})^2 \log(2)}{({\gamma_{s,k}^{(i)}}+1)|\v_{s,k}^{\herm} \H_{k}^{\herm} \w^{(i)}_{s, k}|^{2}}
\end{align}

\noindent and
\begin{align}
\label{eq:mu} \mu_{s,k} & \triangleq \frac{ \zeta_k(1-\alpha) (\bar \gamma_{s,k}^{(i)})^2 \log(2)}{({\bar \gamma_{s,k}^{(i)}}+1) \big|{(\w_{s,k}^{(i)})^{\herm}} \H_{k} \v_{s,k} \big|^{2}}
\end{align}
respectively. Lastly, the dual variables corresponding to~\eqref{eq:dlsinr_cvx_apx} and~\eqref{eq:ulsinr_cvx_apx} are defined as
\begin{align}
\label{eq:nu_bar} \bar \nu_{s,k} & \triangleq \frac{ \eta_k\alpha (\gamma_{s,k}^{(i)})^2 \log(2)}{({\gamma_{s,k}^{(i)}}+1) \big|{(\v_{s,k}^{(i)})^{\herm}} \H_{k}^{\herm} \w_{s, k} \big|^{2}}, \\
\label{eq:mu_bar} \bar \mu_{s,k} & \triangleq \frac{ \zeta_k(1-\alpha) (\bar \gamma_{s,k}^{(i)})^2 \log(2)}{({\bar \gamma_{s,k}^{(i)}}+1)| \w^{\herm}_{s, k} \H_{k} \v_{s,k}^{{(i)}}|^{2}}
\end{align}
respectively.

\bibliographystyle{IEEEtran}
\bibliography{IEEEabbr,refs}
\end{document}

%% file: Definitions.tex
\newcommand{\h}{\mathbf{h}}

\newcommand{\p}{\mathbf{p}}

\renewcommand{\v}{\mathbf{v}}
\newcommand{\w}{\mathbf{w}}

\newcommand{\E}{\mathbf{E}}

\renewcommand{\H}{\mathbf{H}}
\newcommand{\I}{\mathbf{I}}

\newcommand{\X}{\mathbf{X}}
\newcommand{\Y}{\mathbf{Y}}
\newcommand{\Z}{\mathbf{Z}}

\newcommand{\setB}{\mathcal{B}}
\newcommand{\setC}{\mathcal{C}}

\newcommand{\setK}{\mathcal{K}}

\newcommand{\setN}{\mathcal{N}}

\newcommand{\Real}{\mbox{$\mathbb{R}$}}
\newcommand{\Compl}{\mbox{$\mathbb{C}$}}

\newcommand{\herm}{\mathrm{H}}

\newcommand{\tran}{\mathrm{T}}

%% file: cmnMinFairness.bbl
\begin{thebibliography}{10}
\providecommand{\url}[1]{#1}
\csname url@samestyle\endcsname
\providecommand{\newblock}{\relax}
\providecommand{\bibinfo}[2]{#2}
\providecommand{\BIBentrySTDinterwordspacing}{\spaceskip=0pt\relax}
\providecommand{\BIBentryALTinterwordstretchfactor}{4}
\providecommand{\BIBentryALTinterwordspacing}{\spaceskip=\fontdimen2\font plus
\BIBentryALTinterwordstretchfactor\fontdimen3\font minus
  \fontdimen4\font\relax}
\providecommand{\BIBforeignlanguage}[2]{{%
\expandafter\ifx\csname l@#1\endcsname\relax
\typeout{** WARNING: IEEEtran.bst: No hyphenation pattern has been}%
\typeout{** loaded for the language `#1'. Using the pattern for}%
\typeout{** the default language instead.}%
\else
\language=\csname l@#1\endcsname
\fi
#2}}
\providecommand{\BIBdecl}{\relax}
\BIBdecl

\bibitem{Atz21}
I.~Atzeni, B.~Gouda, and A.~T\"{o}lli, ``Distributed precoding design via
  over-the-air signaling for cell-free massive {MIMO},'' \emph{{IEEE} Trans.
  Wireless Commun.}, vol.~20, no.~2, pp. 1201--1216, Feb. 2021.

\bibitem{Gou20}
B.~Gouda, I.~Atzeni, and A.~T\"{o}lli, ``Distributed precoding design for
  cell-free massive {MIMO} systems,'' in \emph{Proc. {IEEE} Int. Workshop
  Signal Process. Adv. in Wireless Commun. (SPAWC)}, May 2020.

\bibitem{Atz20}
I.~Atzeni, B.~Gouda, and A.~T\"{o}lli, ``Distributed joint receiver design for
  uplink cell-free massive {MIMO},'' in \emph{Proc. {IEEE} Int. Conf. Commun.
  (ICC)}, June 2020.

\bibitem{Zha20}
J.~Zhang, E.~Björnson, M.~Matthaiou, D.~W.~K. Ng, H.~Yang, and D.~J. Love,
  ``Prospective multiple antenna technologies for beyond 5{G},'' \emph{{IEEE}
  J. Sel. Areas Commun.}, vol.~38, no.~8, pp. 1637--1660, Aug. 2020.

\bibitem{Buz19}
S.~Buzzi, C.~{D'Andrea}, A.~Zappone, and C.~{D'Elia}, ``User-centric {5G}
  cellular networks: Resource allocation and comparison with the cell-free
  massive {MIMO} approach,'' \emph{{IEEE} Trans. Wireless Commun.}, vol.~19,
  no.~2, pp. 1250--1264, Feb. 2019.

\bibitem{Int19}
E.~Interdonato, G.~Björnson, H.-Q. Ngo, P.~Frenger, and E.~G. Larsson,
  ``Ubiquitous cell-free massive {MIMO} communications,'' \emph{EURASIP J.
  Wireless Commun. Netw.}, vol. 2019, no.~1, pp. 197--209, Aug. 2019.

\bibitem{Raj20}
\BIBentryALTinterwordspacing
N.~{Rajatheva}, I.~{Atzeni}, E.~{Björnson} \emph{et~al.}, ``White paper on
  broadband connectivity in {6G},'' June 2020. [Online]. Available:
  \url{http://jultika.oulu.fi/files/isbn9789526226798.pdf}
\BIBentrySTDinterwordspacing

\bibitem{Ngo17}
H.-Q. Ngo, A.~Ashikhmin, H.~Yang, E.~G. Larsson, and T.~L. Marzetta,
  ``Cell-free massive {MIMO} versus small cells,'' \emph{{IEEE} Trans. Wireless
  Commun.}, vol.~16, no.~3, pp. 1834--1850, Mar. 2017.

\bibitem{Bjo20}
E.~Björnson and L.~Sanguinetti, ``Making cell-free massive {MIMO} competitive
  with {MMSE} processing and centralized implementation,'' \emph{{IEEE} Trans.
  Wireless Commun.}, vol.~19, no.~1, pp. 77--90, Jan. 2020.

\bibitem{Gou21}
B.~Gouda and A.~T\"{o}lli, ``Distributed joint receiver design for cell-free
  massive {MIMO} with fast convergence,'' in \emph{Proc. {IEEE} Int. Conf.
  Commun. (ICC)}, June 2021.

\bibitem{Tol19}
A.~Tölli, H.~Ghauch, J.~Kaleva, P.~Komulainen, M.~Bengtsson, M.~Skoglund,
  M.~Honig, E.~Lahetkangas, E.~Tiirola, and K.~Pajukoski, ``Distributed
  coordinated transmission with forward-backward training for {5G} radio
  access,'' \emph{{IEEE} Commun. Mag.}, vol.~57, no.~1, pp. 58--64, Jan. 2019.

\bibitem{Bas19}
M.~Bashar, K.~Cumanan, A.~G. Burr, M.~Debbah, and H.-Q. Ngo, ``On the uplink
  max–min {SINR} of cell-free massive {MIMO} systems,'' \emph{{IEEE} Trans.
  Wireless Commun.}, vol.~18, no.~4, pp. 2021--2036, Apr. 2019.

\bibitem{Bas19a}
M.~Bashar, K.~Cumanan, A.~G. Burr, H.-Q. Ngo, E.~G. Larsson, and P.~Xiao,
  ``Energy efficiency of the cell-free massive {MIMO} uplink with optimal
  uniform quantization,'' \emph{{IEEE} Trans. Green Commun. and Netw.}, vol.~3,
  no.~4, pp. 971--987, Dec. 2019.

\bibitem{Zho20}
A.~Zhou, J.~Wu, E.~G. Larsson, and P.~Fan, ``Max-min optimal beamforming for
  cell-free massive {MIMO},'' \emph{{IEEE} Commun. Lett.}, vol.~24, no.~10, pp.
  2344--2348, Oct. 2020.

\bibitem{Scu17}
G.~Scutari, F.~Facchinei, and L.~Lampariello, ``Parallel and distributed
  methods for constrained nonconvex optimization---{P}art {I}: Theory,''
  \emph{{IEEE} Trans. Signal Process.}, vol.~65, no.~8, pp. 1929--1944, Apr.
  2017.

\bibitem{Lee15}
J.~Lee, J.~Liang, J.~Park, and M.~Kim, ``Directional path loss characteristics
  of large indoor environments with 28 {GHz} measurements,'' in \emph{Proc.
  {IEEE} Int. Symp. Pers., Indoor and Mobile Radio Commun. (PIMRC)}, Aug. 2015.

\end{thebibliography}
